\documentclass[aps,prl,groupedaddress,showpacs,latterpaper,preprintnumbers,amsmath,amssymb,twocolumn]{revtex4}
\usepackage{graphicx}

\begin{document}
\title{Effect of amplification on conductance distribution of a disordered waveguide}
\author{Alexey Yamilov$^{1,2}$\footnote{AY is currently with University of Missouri--Rolla} and Hui Cao$^{1}$}
\affiliation{
$^{1}$Department of Physics and Astronomy, Northwestern University, Evanston, IL, 60208\\
$^{2}$Department of Physics, University of Missouri--Rolla, Rolla, MO 65409}
\date{\today}

\begin{abstract}
Introduction of optical gain to a disordered system results in enhanced fluctuations [$F_{(2)}=var(\tilde{g})/\langle \tilde{g} \rangle^2$] of dimensionless conductance $\tilde{g}$, similar to the effect of Anderson localization in passive medium. Using numerical simulations we demonstrate that despite of such qualitative similarity, the whole distribution of conductance of amplifying random media is drastically different from that of passive system with the same value of $F_{(2)}$. 
\end{abstract}
\pacs{42.25.Dd,42.25.Bs,42.55.Zz}


\maketitle

Wave transport in quasi-1D (Q1D) is a paradigm of mesoscopic physics\cite{mello_book}. Universality of the statistics of wave transport in Q1D geometry makes it a convenient system to study interference effects. Anderson localization phenomenon\cite{anderson} is an ultimate manifestation of the wave interference\cite{altshuler,ping_sheng}. Light transport in a Q1D system is described by three directly measurable quantities\cite{feng_and_vanrossum}: $T_{ab}$ -- the transmission coefficient from an incoming channel $a$ to an outgoing channel $b$; $T_a=\sum\limits_b{T_{ab}}$ -- the total transmission coefficient from channel $a$ to all outgoing channels; $\tilde{g}=\sum\limits_{a,b}{T_{ab}}$ -- conductance. This is unlike electronic systems where only conductance can be accessed experimentally. Landauer formula\cite{landauer} puts the transport of electronic\cite{altshuler} and electromagnetic waves\cite{ping_sheng} on the same footing. Average value of conductance ($\langle \tilde{g}\rangle \equiv g\gg 1$ -- diffusion, $g\ll 1$ -- localization), its variance, and more generally, its entire distribution has been found to be indicative of the nature of wave transport through random medium\cite{beenakker_rmp,mirlin}.  Coherent amplification of light adds a new dimension to the study of mesoscopic transport that can lead to new phenomena such as random lasing\cite{cao_review}. 

The effect of amplification leads to enhancement of nonlocal correlations\cite{corr_active} and the fluctuations of various transport coefficients\cite{zyuzinPRE,pofi} that bears similarity to that of localization. Selective amplification of the quasi-modes with long lifetime increases their contribution to transport, leading to ``localization by gain'' \cite{corr_active,pofi,disphc,cao_JPA}. In our previous work we found that $P(T_{ab})$ in active random medium coincides\cite{pofi}, within numerical accuracy, with that of a passive system\cite{kogan_first} with reduced value of dimensionless conductance which parameterizes the distribution. The question arises about its relevance to the actual value of conductance and its distribution in random medium with gain.

Scaling theory of localization\cite{gangof4} predicts that distribution of dimensionless conductance, $\tilde{g}$, depends on a single parameter. The average value $\langle \tilde{g}\rangle$ is usually chosen for the parameterization\cite{beenakker_rmp,mirlin}. Q1D geometry (optical waveguide) is a particularly convenient system to study localization-delocalization transition, which can be realized with an variation of the length of waveguide. 

Introduction of optical gain to the system increases the average conductance of the waveguide. At the same time, it is also expected to enhance the conductance fluctuations. It has been shown that fluctuations of both $T_{ab}$\cite{corr_active} and $T_a$\cite{zyuzinPRE} grow faster than the corresponding averages. However, the effect of gain on statistical properties of $\tilde{g}$ is not known. 

In Ref. \cite{corr_active} in order to obtain correlation functions that contain meaningful information about light transport in amplifying random media, we introduced the conditional average over all non-lasing realizations $\langle ...\rangle \rightarrow \langle ...\rangle_c$. Due to fluctuations of lasing threshold such realizations always exist however improbable. This fact makes analytical treatment of the  problem difficult.

In this work, following the same approach\cite{corr_passive,corr_active}, we obtained numerically the full distribution of conductance in passive and active disordered waveguides with time reversal symmetry\cite{beenakker_rmp}. We demonstrate that the second moment $F_{(2)}\equiv var(\tilde{g})/g^2 = \left( \langle \tilde{g}^2 \rangle - g^2  \right) / g^2$ increases with gain and that the conductance distribution $P(\tilde{g})$ in amplifying system is drastically different from the distribution of passive system with the same second moment $F_{(2)}$.

\begin{figure}
\centerline{\rotatebox{0}{\scalebox{0.35}{\includegraphics{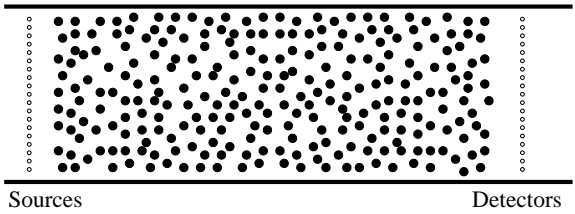}}}}
\caption{\label{geometry} Disordered waveguide geometry.}
\end{figure}

In our numerical simulation, we consider 2D waveguide filled with random medium (Fig. \ref{geometry}). The walls of the waveguide are metallic, and circular scattering particles are dielectric with refractive index $n=2$. We use finite difference time domain method to calculate the response of our system to pulsed excitation, followed by Fourier transformation which gives us the desired continuous-wave response\cite{corr_passive}. We have successfully used this method to study mesoscopic fluctuations and nonlocal correlations\cite{corr_active,pofi}. The system we are considering is Q1D, and the transition from diffusion to localization can be realized by increasing the length $L$ of the random medium. The effect of absorption or gain (inside the scatterers) is treated by classical Lorenzian model with positive or negative conductivity. The advantage of our numerical model is the ability to introduce spatially uniform gain as well as to separate coherent amplification of an input signal from spontaneous emission of the active medium. In the presence of gain, long after the short excitation pulse, the electromagnetic field decays with time in the non-lasing realizations, while it keeps increasing in the lasing ones.  We excluded the lasing realizations from our statistical ensemble.

In order to calculate dimensionless conductance, $\tilde{g}$, a short pulse was launched via a series of equidistant point sources which uniformly cover the input end of the waveguide. One calculation was performed for each source in the sequence, and the transmission coefficient was obtained in the series of detector points which uniformly cover the output end of the waveguide. The summation of  $\tilde{T}_{ab}$ over source and detector points gave us a quantity which is proportional to dimensionless conductance, $\sum\limits_{a,b}{\tilde{T}_{ab}}=\alpha\tilde{g}$. The coefficient of proportionality $\alpha$ is geometrical, and is constant for all realizations. It can be found as follows\cite{my_conductance}. According to scaling theory of localization \cite{gangof4}, the distribution of conductance in a passive system is uniquely determined by its average value. We use normalization-free second moment $F_{(2)}$ of the system to find actual value of average conductance $g$. This can be done using results of Ref. \cite{mirlin}, where $g$ and $var(\tilde{g})$ were calculated analytically. Note that once the coefficient $\alpha$ is found from passive system (where $F_{(2)}$ and $g$ are uniquely related) it can be used in the amplifying or absorbing systems with the same arrangement of the source and detector points.

\begin{figure}
\vskip -0.7cm
\centerline{\rotatebox{0}{\scalebox{0.45}{\includegraphics{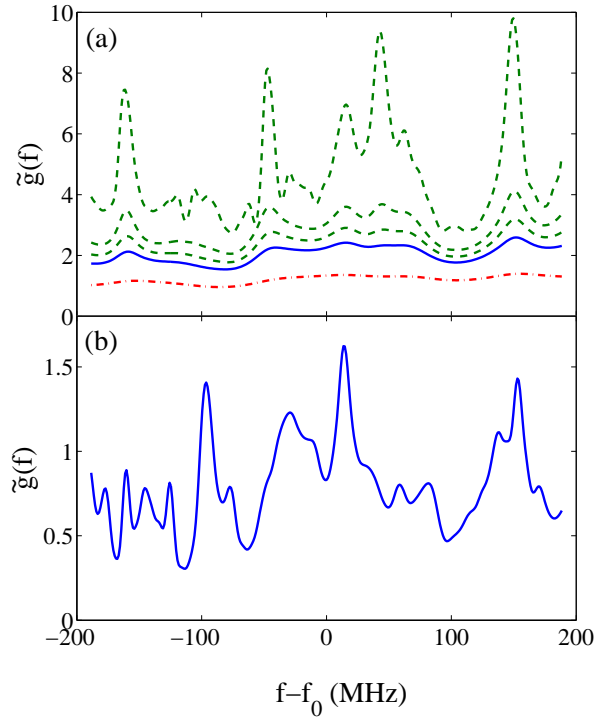}}}}
\vskip -0.7cm
\caption{\label{g_versus_omega} (a) Solid, dash-dotted and dashed lines represent $\tilde{g}(f)$ in passive, absorbing and amplifying systems respectively.  Three dashed curves (from bottom to top) correspond to the gain coefficient $\sigma_g/ \sigma_{g,cr} = 0.11,\ 0.22$ and $0.33$ ($\sigma_{g,cr}$ is the critical value for average lasing threshold \cite{corr_active}). The absorption coefficient for the dash-dotted line $\sigma_{a} = - 0.45 \sigma_{g,cr}$.  (b) $\tilde{g}(f)$ of a passive system with nearly the same $F_{(2)}$ as the upper dashed line of (a).}
\end{figure}

Far into the diffusive regime $P(\tilde{g})$ has a Gaussian shape with the width determined by the universal conductance fluctuations. In terms of quasi-modes of diffusive system\cite{absorption}, addition of weak uniform gain to the system should result in partial compensation of the modes decay rates. Significant changes should occur when a substantial fraction of the quasi-modes are brought sufficiently close to their lasing threshold. This is the regime which we would like to investigate in our Q1D system, that lead to the following choice of system parameters. We start with the passive system close to the onset of localization with $g=1.53$, as determined by the procedure outlined above. The waveguide supports $N=20$ modes. 

Solid line in Fig. \ref{g_versus_omega}(a) shows frequency dependence\cite{lambda_normalization} of the $\tilde{g}$ in one of random realizations of the passive system. As one adds gain (dashed lines) or absorption (dash-dot line), the conductance increases or decreases respectively. Absorption is known to reduce fluctuations of the conductance\cite{brouwer,azi_nature} and makes the distribution approach Gaussian\cite{markos_pofi}. The latter fact led to a proposition of  statistical formulation of localization criterion\cite{azi_nature} in absorbing systems: parameter $g^\prime =(2/3) var[T_a/\langle T_a\rangle ]$ takes on the role of the average dimensionless conductance $g$ as the localization parameter. It was conjectured that localization in absorbing system should be marked by the reduction of $g^\prime$ to below unity, similar to $g<1$ in passive systems. 

Similarly, we can introduce an alternative localization parameter based on the fluctuation of conductance. Fig. \ref{g_versus_omega}(a) shows that introduction of gain leads to enhanced  fluctuations of conductance despite of the increase of its average value. We define a new  localization parameter $g^{\prime\prime}=(2/15) F_{(2)}$, where the numerical factor $2/15$ came from the leading term in the perturbative expansion of variance of $\tilde{g}$ in the diffusive regime\cite{beenakker_rmp,mirlin}. Based on data from Ref. \cite{mirlin} we estimate that in passive systems $g^{\prime\prime}=1.0$ when $g=0.91$. 

In order to make comparative analysis of the full distribution of conductance in active medium ({\it c.f.} Fig. \ref{g_versus_omega}(a)) with that in passive medium ({\it c.f.} Fig. \ref{g_versus_omega}(b)) we first find a passive system with the same value of second moment $F_{(2)}$. Then we scale down the conductance of system with gain $\tilde{g}_s=C\tilde{g}$ so that $\langle\tilde{g}_s\rangle=g_{passive}$. Note that $F_{(2)}=var(\tilde{g})/g^2$ will not be affected by the scaling.

\begin{figure}
\centerline{\rotatebox{-90}{\scalebox{0.37}{\includegraphics{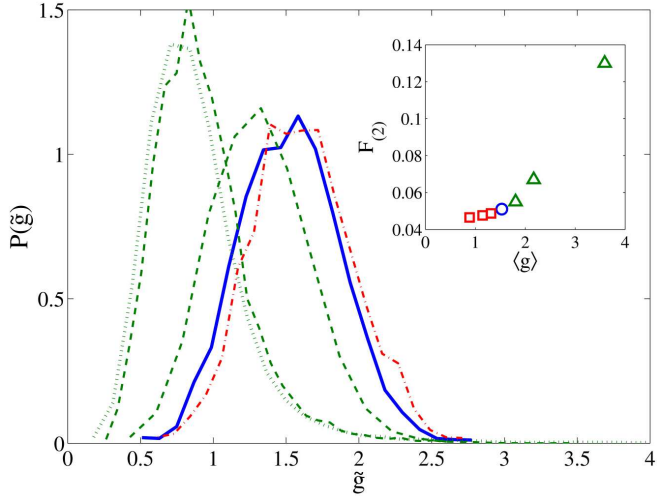}}}}
\caption{\label{p_of_g_gain} The inset plots the normalized second moment $F_{(2)}=var(\tilde{g})/g^2$ as a function of the average conductance $g$ in system with different amount of gain (triangles correspond to $\sigma_g/\sigma_{g,cr}=0.11,\ 0.22$ and $0.45$) or absorption (squares correspond to $\sigma_{abs}/\sigma_{g,cr}=-0.11,\ -0.22$ and $-0.45$). Passive system (circle) is shown for reference. In the main plot, solid, dash-dotted and dashed lines represent $P(\tilde{g})$ in passive ($g=1.53$) or $P_c(\tilde{g}_s)$ in absorbing ($\sigma_a/\tau_{g,cr}=-0.45$) and amplifying (two dashed curves from right to left correspond to $\sigma_g/\sigma_{g,cr}= 0.22$ and $0.45$) systems. Geometrically all systems are the same and differ only by the presence/absence of gain or absorption. The difference between left-most dashed curve and dotted curve is discussed in the text.}
\end{figure}

Figure \ref{p_of_g_gain} shows the change of $P(\tilde{g}_s)$ with introduction of gain (absorption). From the inset of Fig. 3 we see that in amplifying random medium as the optical gain coefficient increases, $var(\tilde{g})$ increases faster than the average value of conductance (i.e. normalized second $F_{(2)}$ moment increases with $g$). This is also evident from the shift of the scaled distribution towards smaller conductances. The increase of $F_{(2)}$ results in smaller $\langle \tilde{g}_s\rangle$ by virtue of the scaling. Secondly, the shape of the distribution also changes, which we will discuss below. The effect of absorption is significantly more moderate: the dash-dotted (absorption) and the left dashed curve correspond to systems with the same absorption/gain coefficients. 

\begin{figure}
\centerline{\rotatebox{-90}{\scalebox{0.37}{\includegraphics{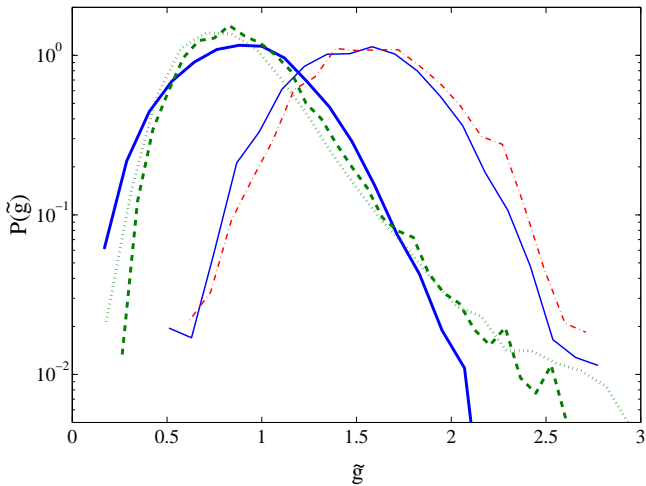}}}}
\caption{\label{p_of_g_compare} Conditional distributions of the scaled dimensionless conductance $\tilde{g}_s$ (dashed and dotted lines) in random amplifying medium ($\sigma_g/\sigma_{g,cr}=0.45$) is compared with that in passive medium (bold solid line) with nearly the same  $F_{(2)}$. Thin solid and dash-dotted lines (same as in Fig. \ref{p_of_g_gain}) correspond to the original passive system and absorbing system with absorption length equal to amplification length of the system with gain.}
\end{figure}

Now we would like to return to the question of definition of conditional statistics in amplifying random media that we first discussed in Ref. \cite{corr_active}. Our numerical algorithm allows calculation of transport coefficients within a certain frequency range. This range is chosen narrow enough so that there is no appreciable change in physical parameters such as transport mean free path\cite{corr_passive}. Therefore, we use the conductance at different frequencies to increase the number of entries in our statistical ensembles to about $\sim 10^{5}$. When obtaining conditional statistical ensemble we dropped the disorder realizations that exceeded the lasing threshold. Thus, the conductances at the non-lasing frequencies for the lasing configurations are dropped, although lasing does not necessarily affect the conductances in the entire frequency range obtained numerically. Direct calculation of conductance in this paper allows us to test the extent of this effect. In Figs. \ref{p_of_g_gain},\ref{p_of_g_compare} we plotted $P_c(\tilde{g}_s)$ calculated using two conditional ensembles. (i) Disorder configurations (and thus contribution from all frequencies in that realization) which lead to diverging intensity are dropped -- dashed lines in the figures. (ii) The entire $P(\tilde{g})$ is plotted first. The obtained distribution contained two separate parts, the second one at large $\tilde{g}$ is artificial and contains the lasing contributions. Its position is determined by the running time of our simulation and therefore can be well separated from the contribution of nonlasing configurations/frequencies. By discarding the contributions above the gap separating two maxima in $P(\tilde{g})$ we obtain the sought conditional distribution -- dotted curves in Figs. \ref{p_of_g_gain},\ref{p_of_g_compare}. 

Comparison of two conditional distributions shows that (ii) leads to slightly greater fluctuations. In case of our samples $F_{(2),i}=0.12$ whereas $F_{(2),ii}=0.13$ compared to $F_{(2)}=0.055$ in the passive system. Removal of the contribution from lasing configurations at non-lasing frequencies reduces fluctuations of conductance and may not be well justified. However, it leads to relatively small correction as it can be seen from Figs. \ref{p_of_g_gain},\ref{p_of_g_compare}.

The question arises: how does $P_c(\tilde{g}_s)$ compares to $P(\tilde{g})$ of passive system (which exhibits the same fluctuation, $F_{(2)}$)? Fig. \ref{p_of_g_compare} compares the conditional distributions of dimensionless conductance in amplifying random medium with that in passive medium with the same $F_{(2)}$  ($g_{passive}=0.92$)\cite{step_feature}. One can see that the distribution in the system with gain has markedly narrower peak, but significantly more extended tail toward large $\tilde{g}_s$. This is in contrast to the intensity distribution\cite{pofi} for which we found a good fit with the formula\cite{kogan_first} of passive system parameterized with the reduced value of conductance. The result presented in Figs. \ref{p_of_g_gain},\ref{p_of_g_compare} suggests that strong modification of conductance distribution occurs in the systems with gain.  

Optical gain is known to enhance the contribution from photons which propagate along long paths and therefore spend long time inside random medium before escaping\cite{wiersma}. These rare long paths should make contributions to the tails of the conductance distribution of passive random medium. This is because long paths are likely to close on themselves and form loops\cite{feng_and_vanrossum}, that is known to lead to universal conductance fluctuations\cite{altshuler}. In the language of quasi-modes (eigenmodes of Maxwell's equations of a finite-size random medium without gain or absorption), with the introduction of optical gain, the quasi-modes with longer lifetime experience more amplification, and they generate the large peaks in the conductance spectrum [Fig. 2(a)]. Thus they form the large-$\tilde{g}_s$ tail of $P_c(g_s)$. Meanwhile, small-$g$ tail should be greatly suppressed. The above conclusions are indeed supported by our numerical data, {\it c.f.} Fig. \ref{p_of_g_compare}. Strong reshaping of the conductance distribution with an increase of amplification also explains why it differs drastically from that of passive random medium with the same variance.

In conclusion, we computed numerically the distribution of dimensionless conductance in the systems with and without optical gain. The fluctuation of conductance is amplified in active random media.  We introduced the concept of conditional distribution of conductance $P_c(\tilde{g}_s)$ that omits the contribution of lasing realizations of random structures and scales the conductance for comparison with that of passive system. Unlike conditional intensity distribution of $P_c(T_{ab})$\cite{pofi}, the obtained conductance distribution can no longer be fitted by that of passive  medium with reduced value of the average conductance. $P_c(\tilde{g}_s)$ in an amplifying random medium differs significantly from that of the passive system with the same value of normalized variance, $F_{(2)}=var(\tilde{g})/\langle \tilde{g} \rangle^2$. It is skewed towards large values of $\tilde{g}$ that we attribute to a dramatic reshaping of the distribution of quasi-modes. In passive localized systems the transport is dominated by the modes that show the strongest coupling to the outside baths, whereas in amplifying systems the most decoupled, confined modes would be preferentially amplified and relocated into large-$\tilde{g}$ tail of the conductance distribution.

This work is supported by the National Science Foundation under Grant No. DMR 0093949. AY also acknowledges support from University of Missouri-Rolla. Correspondence should be sent to yamilov@umr.edu.


\begin{thebibliography}{99}
\bibitem{lambda_normalization} The considered problem is based on linear equations, thus the frequency normalization arbitrary. For concreteness, in Fig. \ref{g_versus_omega} we used $\lambda=2cm$.
\bibitem{step_feature} For $\langle \tilde{g}\rangle <1$ the distribution of conductance develops a step feature at $g=1$\cite{markos_pofg}. It occurs, however, at yet smaller $\langle g\rangle$ \cite{markos_pofg,my_conductance}.
\end{thebibliography}
\end{document}